\documentstyle[11pt,newpasp,twoside,epsf]{article}
\def\edcomment#1{\iffalse\marginpar{\raggedright\sl#1\/}\else\relax\fi}
\reversemarginpar

\begin{document}
\def\lapp{\ifmmode\stackrel{<}{_{\sim}}\else$\stackrel{<}{_{\sim}}$\fi}
\def\gapp{\ifmmode\stackrel{>}{_{\sim}}\else$\stackrel{>}{_{\sim}}$\fi}
\title{Two clocks in the PSR J0737--3039 binary system and their implications 
for the system's origin and evolution}
 \author{D.R. Lorimer,$^{1}$ 
M. Burgay,$^{2}$ 
P.C.C. Freire,$^{3}$
A.G. Lyne,$^{1}$ 
M. Kramer,$^{1}$ 
A. Possenti,$^{4}$ 
M.A. McLaughlin,$^{1}$ 
F. Camilo,$^{5}$ 
R.N.~Manchester,$^{6}$ 
N.~D'Amico,$^{4}$ 
B.C. Joshi$^{7}$} 
\affil{$^{1}$University of Manchester, Jodrell Bank Observatory, SK11 9DL, UK\\
$^{2}$Universita degli Studi di Bologna, Dipartimento di Astronomia, Italy\\
$^{3}$NAIC, Arecibo Observatory, HC3 Box 53995, Arecibo, PR 00612, USA\\
$^{4}$INAF - Osservatorio Astronomica di Cagliari, 09012 Capoterra, Italy\\
$^{5}$Columbia Astrophysics Laboratory, 550 W 120th St, NY 10027, USA\\
$^{6}$ATNF, CSIRO, PO Box 76, Epping, NSW 2121, Australia\\
$^{7}$NCRA, PO Box Bag 3, Ganeshkhind, Pune 411007, India}

\begin{abstract}
As discussed elsewhere in these proceedings, the double pulsar system 
is a magnificent laboratory for gravitational physics and for studying
pulsar magnetospheres. Here we consider the uses of having two clocks 
in the system in the context of its origin and evolution.  
We find that the ``standard'' evolutionary scenario involving spin-up of
the first-born neutron star in an X-ray binary phase is consistent with
the observed parameters. Equality of the spin-down ages of the two
pulsars requires that the post-accretion spin period of A most likely
lies in the range $16 \lapp P_{\rm 0,A} \lapp 21$ ms. The likely
age of the system is $30-70$ Myr.
\end{abstract}

\section{Introduction}

An application of the standard binary pulsar evolutionary model 
(e.g.~Bhattacharya \& van den Heuvel 1992) to the double pulsar system
J0737--3039 identifies the 22.7-ms pulsar (hereafter A) as the
first-born neutron star.  Following an initial phase where A existed
as a regular radio pulsar with a main-sequence companion, the
currently observed spin parameters of A are the result of a subsequent
X-ray binary phase where it acquired matter and angular momentum from
the secondary star after the secondary evolved off the main sequence
and overflowed its Roche lobe.  By processes that are not fully
understood, this mass-transfer phase also resulted in a reduction of
A's magnetic field (Shibazaki et al.~1989).  Following the
supernova explosion of the secondary, a second neutron star was formed
which we now observe as the 2.77-s pulsar (hereafter B).

While the spin periods of both these pulsars, and the low inferred
magnetic field of A relative to B\footnote{Due to the interaction of
A's wind which penetrates deep into B's magnetosphere, some care
should be taken when interpreting B's magnetic field strength.  In
spite of this disruption, B is observed to be 
spinning down due to a steady braking
torque which we model in Section 2.} are consistent with this
model, a further test is to compare the time since the spin-up
phase of A ended (${\cal T}_{\rm A}$) with the time since B has been
active as a radio pulsar (${\cal T}_{\rm B}$). We therefore
expect ${\cal T}_{\rm A}={\cal T}_{\rm B}$.

The simplest means to test this prediction is to use the
characteristic ages of A and B, which are based on the observed
periods and period derivatives: $\tau_{\rm A}=P_{\rm A}/(2
\dot{P}_{\rm A})$ and $\tau_{\rm B}=P_{\rm B}/(2 \dot{P}_{\rm B})$.
Lyne et al.~(2004) find $\tau_{\rm A}=2.1 \times
10^8$ yr and $\tau_{\rm B}=0.5 \times 10^8$ yr.  Possible explanations
for this discrepancy are: (i) the standard evolutionary scenario does
not apply;
(ii) as observed in other pulsars (see e.g.~Kramer et
al.~2003), characteristic ages are not reliable; (iii) both
the model and the characteristic ages are wrong!  Given the
aforementioned circumstantial evidence in favour of recycling hypothesis,
the simplest solution is option
(ii). We now briefly investigate the implications for this case.
Further details will be given in a forthcoming paper
(Lorimer et al.~in preparation).

\section{Modeling the spin-down for A and B}

We consider a generic pulsar spin-down model of the form
\begin{equation}
\dot{\Omega} = K \Omega^n,
\end{equation}
where for a spin period $P$, the angular frequency $\Omega=2\pi/P$,
$n$ is the braking index (for spin-down due to magnetic dipole
radiation, $n=3$) and $K$ depends on the braking torque applied to the
star.  In the simplest case, both $K$ and $n$ are independent of time
and equation (1) can be integrated directly. For the case $n \neq 1$,
we find the ``true age'' of the pulsar
\begin{equation}
t_{\rm true} = 
\frac{2 \tau}{(n-1)} \left[1-\left(\frac{P_0}{P}\right)^{n-1}\right],
\end{equation}
where $P_0$ is the initial spin period and $\tau=P/2\dot{P}$ is the
characteristic age. Alternatively, if $K$ decays exponentially with
$1/e$ time scale $t_{\rm decay}$, then
\begin{equation}
t_{\rm reduced} = t_{\rm decay} \ln(1 + t_{\rm true}/t_{\rm decay})
\end{equation}
is the so-called ``reduced age''.  We now consider various
applications of these solutions by simply equating the derived
ages for the two pulsars and determining their resulting initial
spins and ages. In order to distinguish between both pulsars, we
use ``A'' and ``B'' subscripts where appropriate.

\noindent{\bf Case 0: spin-down due to a non-decaying magnetic dipole.}
We first assume that both pulsars spin down due to magnetic dipole
radiation (i.e.~$n_{\rm A}=n_{\rm B}=3$), their braking torques do not
decay (i.e.~$K_{\rm A}$ and $K_{\rm B}$ are constant) and that the
initial spin period of B was much smaller than currently observed
(i.e.~$P_{\rm 0,B} \ll P_{\rm B}$).  Requiring that $t_{\rm
true,A}=t_{\rm true,B}$, the initial spin period of A after the
spin-up phase ($P_{\rm 0,A}$) and the time since spin-up ceased
(${\cal T}$) are:
\begin{equation}
P_{\rm 0,A} = P_{\rm A} \sqrt{1-(\tau_{\rm B}/\tau_{\rm A})}
\simeq 20 \,\mbox{ms}, 
\,\,\,\mbox{and}\,\,\,{\cal T}={\cal T}_{\rm B}=\tau_{\rm B}
\simeq 50\,\mbox{Myr}.
\end{equation}
As mentioned by Lyne et al.~(2004), the value for $P_{\rm 0,A}$
is consistent with the period of J0737--3039A predicted
by spin-up models (e.g.~Arzoumanian et al.~1999).

\begin{figure}[ht!]
\plotone{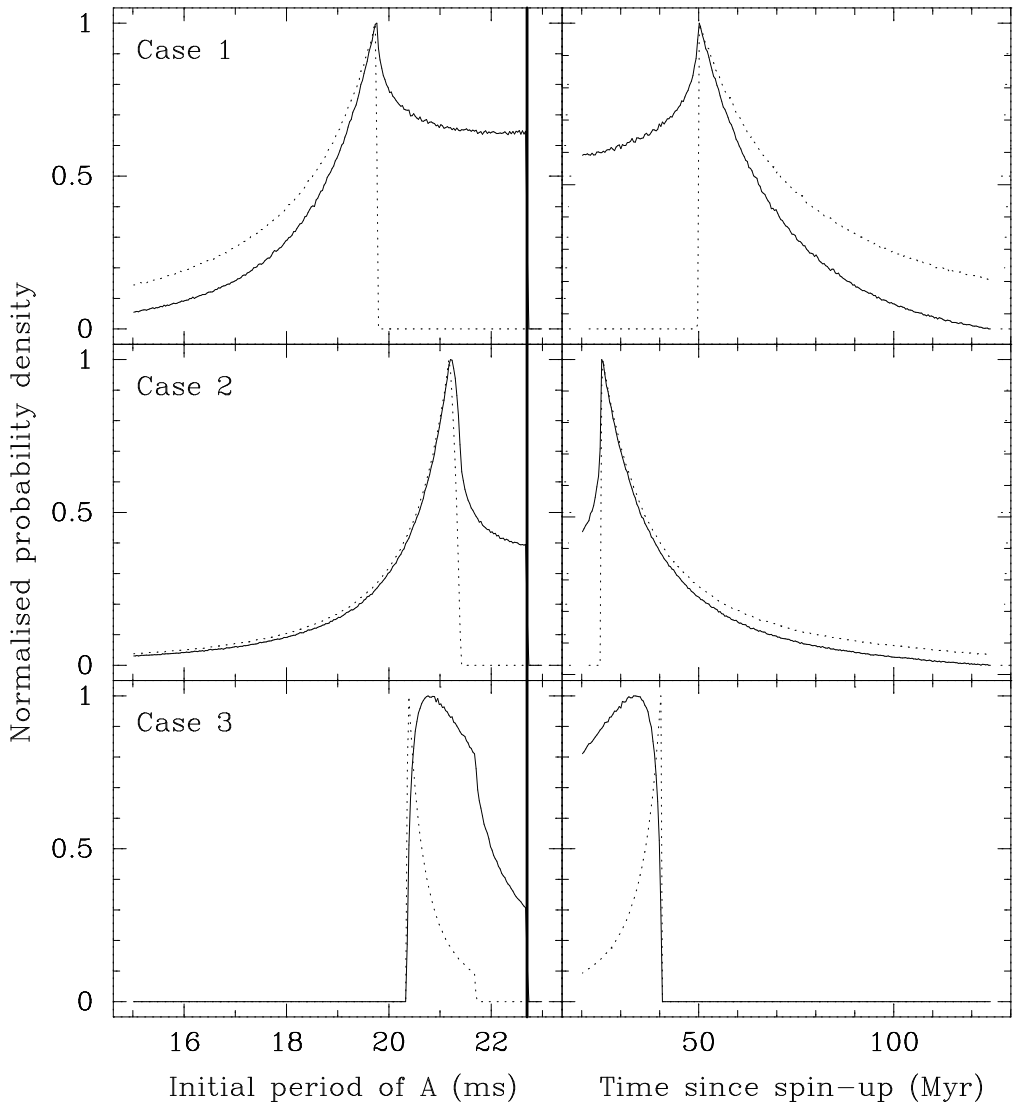}
\caption{Simulated distributions of the post-accretion spin period of
A (left panels) and the time since the end of the spin-up phase (right
panels). In each case, the distributions are calculated assuming a
range of initial spin periods for B (solid curves) and a negligible
initial period (dotted curves).  
The thick vertical line shows A's current spin period.}
\end{figure}

\noindent{\bf Case 1: no torque decay, $n_{\rm A}=3$ and 
$1.4<n_{\rm B}<3.0$.}
A more realistic model is to relax the assumption of dipolar spin-down
for B and allow its braking index to vary in the range observed for
other non-recycled pulsars ($1.4<n_{\rm B}<3.0$; see e.g. Kaspi \&
Helfand 2002 for a review).  For a flat braking index distribution in
this range, a simple Monte Carlo simulation to compute $t_{\rm
true,B}$ for a large number of trials results in the distributions of
$P_{\rm 0,A}$ and ${\cal T}$ for the condition $t_{\rm true,A}=t_{\rm
true,B}$ shown in Fig.~1. In order to show the negligible effect of
B's unknown initial spin period on the results, we performed all
calculations with a flat distribution in the range $0<P_{\rm
0,B}<P_{\rm B}$ (solid lines) and for $P_{\rm 0,B}=0$ (dashed
lines). In both cases, the resulting initial spin period distribution
for A peak sharply just below 20 ms. The age distribution peaks
at $\sim 50$ Myr (i.e.~$\tau_{\rm B}$).

\noindent{\bf Case 2: no torque decay, variable braking indices for A and B.}
Relaxing the conditions on braking indices imposed in case 1,
the centre panel of Fig.~1 shows the results when
$n_{\rm A}$ and $n_{\rm B}$ are drawn from flat distributions in the
range $0 < n < 5$. Regardless of B's initial spin period,
the peak of the $P_{\rm 0,A}$ distribution
is increased over case 1; the age distribution favours smaller 
ages than case 1.

\noindent{\bf Case 3: $n_{\rm A}=n_{\rm B}=3$, exponential torque decay for B.}
The above two cases assume no decay of the braking torque.  An
alternative solution to the spin-down model is the case of an
exponentially decaying braking torque resulting in equation (3)
above. The cause of torque decay is uncertain and controversial, and
thought to be due to either the decay of the neutron star's magnetic
field and/or the alignment of the spin and magnetic axes with time
(see e.g.~Tauris and Konar 2001).  Since torque decay is
not thought to be significant for recycled pulsars (Bhattacharya \&
van den Heuvel 1992), we consider here the case 
in which only the torque on B decays. In Fig.~1 we show
the simulated distributions resulting from the equality
$t_{\rm true,A}=t_{\rm reduced,B}$ assuming pure magnetic dipole braking
($n_A=n_B=3$) and a torque decay on B with a timescale $t_{\rm d,B}$
drawn from a flat distribution between 10 and 100 Myr. The effect
of such a decay is to decrease the age significantly so that
the distribution peaks at around 30-40 Myr.

\section{Conclusions}

The two clocks in the double pulsar system J0737--3039 provide
a unique means to constrain the age and birth spin periods of
the two pulsars.
These simple case studies demonstrate the use of the recycling model
to place constraints on the post-accretion spin period of A ($P_{\rm
0,A}$) and the age since spin-up ceased (${\cal T}$).  For the
spin-down models considered, we find $P_{\rm 0,A}$ to be in the range
$16-21$ ms and ${\cal T}$ to be in the range $30-70$ Myr.  This is
shorter than the age of 100 Myr assumed in the merger
rate calculations by Burgay et al.~(2003) and Kalogera et al.~(2004)
and hence increases the overall neutron star merger rate by 20--40\%.

\end{document}